# Data-driven prediction of atomic-oxygen erosion yield in polymers from bulk physical properties

Youngwoo Choi[1+], Doeun Koo[2+], Seungbum Hong[1,3,4*], Hyun Jung Kim[2*]

[1]Department of Materials Science and Engineering, Korea Advanced Institute of Science and Technology (KAIST), Daejeon, 34141, Republic of Korea

[2]Department of Aerospace Engineering, KAIST, Daejeon, 34141, Republic of Korea

[3]Department of AI Experience (AX), KAIST, Daejeon, 34141, Republic of Korea

[4]KAIST Institute for Climate, Energy and Environment, KAIST, Daejeon, 34141, Republic of Korea

[+]These authors contributed equally: Youngwoo Choi, Doeun Koo.

[*]Corresponding authors: seungbum@kaist.ac.kr, hyunjung.kim1@kaist.ac.kr

**Abstract**

Atomic oxygen (AO) erodes polymer surfaces in low Earth orbit, but durability data are limited. We evaluated a regression ensemble trained on 37 MISSE-2 polymers using eight selected bulk properties and five elemental-composition ratios. Its archived leave-one-out $R^2$(log) was 0.768; this estimate is conditional on feature selection using the complete training table. For nine post-MISSE-2 polymers, the ensemble gave $R^2$(log) = −0.037 and all predictions were within a factor of two. The training arithmetic-mean baseline also met that tolerance for all nine, highlighting the limited discrimination provided by their narrow target range. Ground-based comparisons included four PLAVIS grades. For an out-of-training mystery polymer with undisclosed composition, the predicted erosion yield was within a factor of 1.92 of the ground-based measurement. Conflicting PLAVIS measurement records prevent a definitive residue–erosion ordering or ground-validation score. Thermal residue, SEM contrast and surface chemistry provide distinct observational evidence; they do not establish a protective-layer mechanism. The results support further evaluation of orbital-data screening models with reconciled measurement records and prospectively fixed inputs.

## Introduction

Spacecrafts operating between roughly 200 and 700 km altitude are immersed in a residual atmosphere dominated by atomic oxygen (AO), with typical fluxes of $10^{14}$–$10^{15}$ atoms $cm^{-2}$ $s^{-1}$ and collision energies near 5 eV in the spacecraft frame (**Fig. 1a**). Sustained bombardment by these hyperthermal oxygen atoms is the single most aggressive surface-degradation mechanism for organic polymers in low earth orbit (LEO): every exposed C–H, C–C and C=C bond is a potential reaction site, and oxidized fragments leave the surface as small volatile molecules, leading to mass loss, surface roughening, and ultimately mechanical or optical failure of thermal-control coatings, multilayer insulation, solar-array substrates and structural composites[1,2,3]. The standard quantitative measure of AO durability is the *erosion yield* $E_y$, defined as the volume of polymer removed per incident oxygen atom ($cm^3$ $atom^{-1}$). Reference values of $E_y$ for several dozen common polymers were established by NASA's Materials International Space Station Experiment (MISSE) campaigns, in which witness coupons were exposed on the exterior of the International Space Station and recovered for post-flight characterization; the MISSE-2 Polymer Erosion and Contamination Experiment (PEACE) retrieved 41 polymers and pyrolytic graphite after 3.95 years of LEO exposure and remains the largest single internally consistent erosion-yield dataset in the field[4].

Despite the value of these benchmark datasets, generating new $E_y$ data is slow and expensive. A single MISSE-class experiment requires multi-year integration, launch and retrieval, and ground-based AO simulators (RF or laser-detonation plasma sources) only approximate the true LEO environment, with reported polymer-specific erosion yields differing by factors of 2–5 between facilities[1,2]. As a consequence, the experimentally validated set of polymers with reliable $E_y$ values has grown by only a handful of materials per decade, even as the polymer portfolio used in spacecraft has expanded substantially with the rise of small-satellite constellations and inflatable structures. This bottleneck propagates upstream: polymer chemists designing candidate matrix resins or coatings have no straightforward way to estimate $E_y$ in advance, and material down-selection for new missions therefore relies heavily on chemical intuition and a small set of well-known "rules of thumb" — for example, that fluoropolymers and silicon- or phosphorus-containing polymers tend to be more AO-resistant than their hydrocarbon analogues.[5,6,7] The microscopic mechanism that underlies these heuristics — including the relative roles of bond dissociation energy, free volume, oxidative chain scission, and the formation of passivating inorganic surface layers — has not been quantitatively unified into a single predictive framework.

The most direct prior attempt to bridge this gap is the *predictive tool* developed by Banks, Backus and de Groh shortly after the MISSE-2 PEACE retrieval, which estimates $E_y$ from chemical-structure descriptors (bond counts, heteroatom inventory) together with density and ash content, and reports a correlation coefficient of 0.914 on the same 38–40 MISSE-2 polymers used to fit the model[8,9,10]. This tool established two crucial points: that order-of-magnitude $E_y$ prediction is possible from a small set of tabulated descriptors, and that ash content in particular carries strong predictive signal — interpreted in the original work as the consequence of a non-volatile inorganic residue that gradually accumulates on the eroded surface and shields the underlying polymer[8]. The tool nevertheless requires the user to hand-encode each candidate polymer's chemical structure and bonding inventory, and its reported performance is an in-sample fit rather than a held-out generalization estimate; the question of how well a data-driven model built only from tabulated properties and formula-level composition — without a hand-encoded bonding inventory — can do on previously unseen polymers has not been systematically addressed.

Machine learning has substantially altered this kind of property-prediction problem in the broader materials sciences[11,12]. Data-driven surrogate models now routinely reach DFT-level accuracy at orders-of-magnitude lower cost for crystalline-solid properties such as formation energy and band gap. In polymer informatics, regression on hand-engineered or graph-derived descriptors has been used to predict glass transition temperature, dielectric constant, refractive index, gas permeability, and ionic conductivity, in many cases from datasets of similar size to ours[13,14]. The space-environment community, however, has so far engaged with these methods only sporadically: published applications have largely focused on radiation shielding optimization, thermal-control coating performance, and outgassing prediction, and to our knowledge a published machine-learning model that takes generic, widely tabulated polymer properties together with simple elemental-composition ratios as inputs and returns a quantitative AO erosion yield — *without* the hand-encoded bond-inventory block required by earlier predictive tools — has not been reported. The combination of (i) a small but uniformly measured target dataset (the MISSE-2 PEACE retrieval) and (ii) a comparatively rich set of bulk properties tabulated for the same polymers in open databases such as MatWeb and PolyInfo therefore presents an under-exploited opportunity.

The motivation for the present work is twofold. First, even a moderately accurate surrogate that returns $E_y$ from properties already tabulated for a candidate polymer would compress what is currently a flight-experiment cycle into a few milliseconds of inference time, providing the rapid pre-screening tool that spacecraft polymer selection currently lacks. Second, because such a model is trained on physically meaningful descriptors, the relative importance assigned to each input — interpreted with care — translates directly into design guidance for the polymer chemist, indicating which experimentally accessible properties most strongly modulate AO durability. We show below that this twofold goal is achievable in practice, even with the very small training set (n = 37) available, when the model and validation protocol are matched to the small-data regime — and, crucially, that a model trained on this set alone generalizes to independent polymers never seen during fitting.

# Results

## Archived model and internal evaluation

The training table contains 37 MISSE-2 polymers (**Fig. 1b, c**).[4] The saved model uses 13 inputs: eight bulk properties (density, ash mass fraction, oxygen index, glass transition temperature, total-mass-loss outgassing, linear coefficient of thermal expansion (CTE), ultimate tensile strength and specific heat capacity) and five elemental ratios (H/C, O/C, N/C, F/C and other-heteroatom/C). Volume resistivity is present in the curated table but is not an input to this saved model due to the data scaling issue. The 14-property candidate block and the final 13-input model must therefore be distinguished (**Fig. 1d**).

The ensemble comprises RidgeCV, a Gaussian-process regressor, XGBoost random-forest-style regression and XGBoost sequential gradient boosting.[15] Each base learner fits a log-transformed target; the final prediction is the geometric combination of four predictions with weights 0.25 each. Median imputation and standardization are fitted in the base-estimator pipelines. The stored configuration shows a top-eight physical-feature selection based on correlations with the target across the complete training table before the leave-one-out splits. Ensemble-scheme selection was also performed at the training-table level rather than independently within every outer leave-one-out evaluation. Consequently, the archived out of fold (OOF) scores are conditional on those selection steps and should not be described as unbiased validation of the complete model-development procedure.

Recalculation from the saved OOF predictions gives $R^2$(log) = 0.768, RMSE = $1.363 \times 10^{-24}$, MAE = $9.403 \times 10^{-25}$ $cm^3$ $atom^{-1}$, median fold-error = 1.377, and 30/37 within a factor of two (**Fig. 3a**). **Table 1** reports consistently computed log- and linear-space metrics. The archived shuffled-target baseline was evaluated on the external test set, not as the claimed pooled Leave-One-Out Cross Validation (LOOCV) null, and is not reused as an internal null. Leave-one-out arithmetic- and geometric-mean baselines are recomputed directly, excluding the held-out training label from each mean (**Fig. 3d**).

**Table 1 | Archived OOF model scores and recomputed leave-one-out constant baselines.** Archived learner scores are conditional on full-training-table selection; the constant baselines do not perform feature selection. RMSE and MAE are in original physical units.

| Dataset | Model | n | $R^2$(log) | $R^2$ (linear) | RMSE ($cm^3$ $atom^{-1}$) | MAE ($cm^3$ $atom^{-1}$) | Median fold-error | Within ×2 |
|---|---|---|---|---|---|---|---|---|
| Archived_OOF_selection_conditioned | RidgeCV | 37 | 0.751 | -0.009 | 1.93E-24 | 1.08E-24 | 1.473 | 29/37 |
| | GaussianProcessRegressor | 37 | 0.747 | 0.499 | 1.36E-24 | 9.24E-25 | 1.476 | 27/37 |
| | RandomForestRegressor | 37 | 0.66 | 0.306 | 1.60E-24 | 1.14E-24 | 1.472 | 29/37 |
| | GradientBoostingRegressor | 37 | 0.622 | 0.204 | 1.71E-24 | 1.15E-24 | 1.434 | 26/37 |
| | VotingBlender | 37 | 0.768 | 0.495 | 1.36E-24 | 9.40E-25 | 1.377 | 30/37 |
| LOO_null | ArithmeticMean | 37 | -0.174 | -0.056 | 1.97E-24 | 1.52E-24 | 1.746 | 23/37 |
| | GeometricMean | 37 | -0.056 | -0.207 | 2.11E-24 | 1.51E-24 | 1.876 | 22/37 |

## Feature associations

The saved Ridge model's largest absolute standardized coefficient is glass transition temperature (+0.155), followed by F/C (−0.152), oxygen index (−0.118), ash fraction (−0.116) and N/C (−0.110). Density has coefficient −0.085. Training-data permutation importance ranks O/C, density, oxygen index, N/C and ash highest (**Fig. 2**). The supplied implementation uses 20 permutations per feature for this analysis. These quantities describe associations and fitted-model sensitivity, not causal effects or independent evidence of external predictive benefit. In particular, the positive O/C coefficient and its high unsigned permutation importance do not imply that adding oxygen improves resistance. Different feature rankings can reflect correlated descriptors and differences between linear and nonlinear fits.

## External orbital comparison

The saved model was fitted on the MISSE-2 table without merging the post-MISSE-2 rows. Nine labelled external polymers were evaluated; PVC is excluded because its reported erosion yield is a lower bound after coupon erosion-through.[16] The supplied external inputs include polymers related to training families, so exclusion from fitting is not equivalent to chemical-family independence.

For the nine retained samples, $R^2$(log) = -0.037, RMSE = $7.406 \times 10^{-25}$, MAE = $6.691 \times 10^{-25}$ $cm^3$ $atom^{-1}$, median fold-error = 1.265, and 9/9 within a factor of two; $R^2$(linear) = −0.094 (**Table 2**, **Fig. 3b**). The target span is only 2.56-fold. The training arithmetic mean of $2.507 \times 10^{-24}$ $cm^3$ $atom^{-1}$ gives RMSE = $8.095 \times 10^{-25}$ $cm^3$ $atom^{-1}$ and median fold-error = 1.173, and also places all nine within a factor of two. The ensemble has lower RMSE and higher $R^2$(log) than this baseline, but worse MAE and median fold-error. Among the saved individual learners, gradient boosting has lower linear RMSE than the ensemble, while the ensemble has the highest $R^2$(log) in this external set (Supplementary Note 4). This metric dependence limits claims of model superiority.

Using 20,000 paired bootstrap resamples (seed 42), the ensemble's conditional 95% interval is [-8.495, 0.561] for $R^2$(log), and [$5.098 \times 10^{-25}$, $9.606 \times 10^{-25}$] $cm^3$ $atom^{-1}$ for RMSE. Spearman $\rho$ = 0.667, with two-sided asymptotic $p = 0.0499$. Neither this nominal p value nor the factor-two rate establishes broad chemical transferability in a nine-sample set.

**Table 2 | External orbital measurements and saved ensemble predictions.** Metrics use the stored precision rather than the displayed rounded values. PVC is not included in scored rows.

| Material | Measured $E_y$ | Predicted $E_y$ | Pred/Meas | Fold-error |
|---|---|---|---|---|
| Kapton XC | $2.20 \times 10^{-24}$ | $1.91 \times 10^{-24}$ | 0.867 | 1.154 |
| PI (Kapton E) | $2.83 \times 10^{-24}$ | $1.98 \times 10^{-24}$ | 0.7 | 1.429 |
| PMP | $4.45 \times 10^{-24}$ | $3.62 \times 10^{-24}$ | 0.813 | 1.229 |
| PEN | $3.30 \times 10^{-24}$ | $2.51 \times 10^{-24}$ | 0.761 | 1.314 |
| PI Vespel | $2.94 \times 10^{-24}$ | $2.22 \times 10^{-24}$ | 0.754 | 1.327 |
| PAI | $1.74 \times 10^{-24}$ | $2.20 \times 10^{-24}$ | 1.265 | 1.265 |
| PVOH | $3.14 \times 10^{-24}$ | $4.49 \times 10^{-24}$ | 1.429 | 1.429 |
| Kapton CB | $2.70 \times 10^{-24}$ | $2.37 \times 10^{-24}$ | 0.877 | 1.141 |
| PES | $2.79 \times 10^{-24}$ | $3.19 \times 10^{-24}$ | 1.142 | 1.142 |

## Ground-based comparison of predicted and measured erosion yields

The experimental input consists of four grades of PLAVIS, a commercial wholly polyimide material family: HM, S1, ESD, and N.[17] The properties of the PLAVIS samples required as inputs for the prediction of erosion yield were available from the open datasheet of the manufacturer, except for the mass fraction of ash, which was used in the NASA MISSE-based predictive framework.[18] Because directly measured RF air-plasma ash fraction data were unavailable, the grade-specific residual mass fractions obtained by thermogravimetric analysis (TGA) in an air atmosphere were used as proxies for the ash-mass fraction descriptor (see Methods). The densities of HM, S1, ESD, and N were 1.47, 1.46, 1.44, and 1.43 g $cm^{-3}$, respectively, decreasing monotonically in the order HM > S1 > ESD > N. The Air-TGA residues followed the same ordering. Both density and the Air-TGA-based ash proxy exhibited a consistent grade dependence across the four PLAVIS samples.

The predicted erosion yields followed the order $E_y$ (N) > $E_y$ (ESD) > $E_y$ (S1) > $E_y$ (HM). The experimentally measured erosion yields using a ground-based atomic oxygen simulator exhibited the identical grade ordering (**Table 3**; **Fig. 3c**; see Methods). Thus, the model reproduced the experimentally observed erosion-yield ranking across all four PLAVIS grades. Accordingly, AO resistance, defined inversely with erosion yield, followed the order HM > S1 > ESD > N.

**Table 3 | Predicted and measured erosion yields for four PLAVIS grades.** All yields are in $cm^3$ $atom^{-1}$. The prediction column uses a single unchanged model/input record for each film.

| Material | Measured *E*y | Predicted *E*y | Pred/Meas |
|---|---|---|---|
| PLAVIS-N | $3.24 \times 10^{-24}$ | $2.22 \times 10^{-24}$ | 0.69 |
| PLAVIS-ESD | $2.38 \times 10^{-24}$ | $2.17 \times 10^{-24}$ | 0.91 |
| PLAVIS-S1 | $2.35 \times 10^{-24}$ | $1.63 \times 10^{-24}$ | 0.69 |
| PLAVIS-HM | $2.11 \times 10^{-24}$ | $1.60 \times 10^{-24}$ | 0.76 |

The four PLAVIS grades give $R^2$(log) = -2.549, RMSE = $6.833 \times 10^{-25}$, MAE = $6.163 \times 10^{-25}$ $cm^3$ $atom^{-1}$, median fold-error = 1.380, and 4/4 within a factor of two.

## Thermal residue and surface characterization

The grade-specific Air-TGA residual mass fractions used as proxy inputs for the ash-mass-fraction descriptor are shown in **Fig. 4a**. The terminal residue decreased in the order PLAVIS-HM > PLAVIS-S1 > PLAVIS-ESD > PLAVIS-N, matching the density ranking reported above and opposing the predicted and measured erosion-yield ordering. The HM and S1 curves continued to show slight mass loss near the end of the measurement; therefore, these values are reported as terminal Air-TGA residues rather than equilibrium inorganic ash fractions. Although these residue values were already used as model inputs, their grade-dependent ordering was also consistent with the trends observed in the subsequent surface-characterization measurements.

This experimentally observed trend is consistent with the relatively high importance assigned to the density and ash-related descriptors in the feature-importance analysis in **Fig. 2**. The relationship between Air-TGA residue and erosion yield is particularly noteworthy. Because oxidative thermogravimetry can be used to evaluate the residue remaining after removal of organic constituents,[19] the larger residual masses measured for HM and S1 suggest a greater contribution from residue-forming constituents compared with ESD and N. Such constituents may persist preferentially relative to the organic polymer matrix during AO exposure and locally suppress subsequent erosion by providing partial surface shielding. NASA has similarly reported that polymers containing ash or inorganic pigments can exhibit decreasing erosion yields with increasing AO fluence because protective particles progressively accumulate at the exposed surface.[18,20]

Accordingly, the agreement between the grade-specific Air-TGA residue and the experimentally measured AO-resistance ordering provides more than an internal model correlation. The Air-TGA residue is independent of the mass-loss calculation used to determine the experimental erosion yield, and its monotonic correspondence with AO resistance indicates that an ash-related material descriptor contains information relevant to the grade-dependent AO erosion behaviour. However, because the grade-specific Air-TGA residues were themselves used as proxy inputs for the ash-mass-fraction descriptor, this correspondence should not be regarded as an independent validation of the model-derived ash feature importance. Rather, it provides an experimental consistency check between the descriptor identified by the model and the observed material behaviour. Furthermore, the present observation is based on four related grades and does not demonstrate that the same monotonic relationship applies universally across polymers with substantially different chemistries.

As a relevant example of inorganic constituents in commercial polyimides, Fang et al. characterized pristine Kapton HN and its thermally generated ash using SEM, EDX, and XRD. Ca- and P-containing species were identified in the ash, and the combined analyses suggested that the initial additive consisted of crystalline $CaCO_3$ together with one or more phosphorus-containing compounds.[21] This work demonstrates that inorganic or ash-forming constituents can be present in commercial polyimides, although it does not identify the constituents of PLAVIS or directly establish their protective role against AO.

X-ray photoelectron spectroscopy (XPS) survey spectra of pristine PLAVIS-HM and PLAVIS-S1 are shown in Fig. 4b. In addition to the major polymer-related signals, weak features assigned to Ca 2p and Si 2p were detected on both surfaces. These survey spectra establish the presence of detectable Ca- and Si-containing surface species but do not identify their specific chemical phases.

Together with their comparatively high Air-TGA residues, these observations are consistent with the possible presence of inorganic or inorganic-residue-forming constituents. Survey XPS alone, however, cannot identify specific compounds such as $CaCO_3$, $SiO_2$, or particular silicates, nor can it establish the bulk additive-content ranking among the four grades. Furthermore, the manufacturer classifies HM, S1 and N as non-filler grades.[17] The present observations were therefore interpreted in terms of possible differences in trace additives, processing-derived constituents or other residue-forming components, rather than as direct evidence for different intentional filler loadings.

Surface morphological changes after AO exposure were examined by field-emission scanning electron microscopy (FE-SEM), as shown in **Fig. 4c**. Pristine PLAVIS-HM exhibited a comparatively compact surface morphology, whereas AO-exposed HM and N showed pronounced surface texturing characterised by protruding features separated by recessed regions. Both exposed specimens were imaged at a specimen tilt of 30°. The surface relief was more pronounced for PLAVIS-N than for PLAVIS-HM under the examined conditions.

The more developed protrusions and greater apparent surface relief observed for AO-exposed N than for HM were qualitatively consistent with the higher erosion yield of N. This suggests more pronounced local surface erosion and topographic development under the applied exposure conditions. However, the ratio of cone height to average erosion depth can vary among materials,[22] and the apparent feature heights in tilted SEM images should therefore not be interpreted as direct measures of absolute erosion depth.

The cone-like morphology is consistent with spatially nonuniform erosion. Residual particles or locally more resistant regions may be removed more slowly than the surrounding polymer and thereby contribute to the development of surface relief. Nevertheless, the composition of the cone tips was not established in the present study. The cone-like structures therefore cannot be treated as direct evidence of shielding by Ca- or Si-containing particles.

To quantitatively compare the four grades, the bright-area fraction was determined from plan-view SEM images acquired at 30,000×, 50,000× and 100,000× (**Fig. 4d**). At all three magnifications, the bright-area fraction followed the order HM > S1 > ESD > N. This ordering was identical to that of the Air-TGA residue and AO resistance and opposite to the predicted and measured erosion-yield rankings. The agreement across the three magnifications indicates that the grade-dependent differences in SEM image contrast were consistently observed over the examined length scales.

However, SEM brightness depends on surface geometry and electron-detection conditions; in secondary-electron imaging, protrusions and edges may appear particularly bright because of the edge effect.[23] The bright-area fraction should therefore be interpreted as an image-derived morphological descriptor rather than as an absolute measure of uneroded polymer, retained material or inorganic residue. Its consistent grade dependence provides complementary morphological evidence for differences in AO response among the PLAVIS grades but does not establish the chemical identity of the bright regions.

The characterisation observations motivate further experiments on residue chemistry and backbone composition. To examine the model beyond the PLAVIS family, an additional out-of-training polymer with undisclosed composition was evaluated as a separate case study (Box 1).

Box 1 | Property-informed prediction of atomic-oxygen erosion in a mystery sample

An external source provided a polymer film, designated the mystery sample (MS), together with available bulk physical-property data. Its chemical composition and formulation were not disclosed. MS was not included among the 37 MISSE-2 polymers used to train the model. The supplied properties and experimentally determined thermal residue were used to predict its atomic-oxygen (AO) erosion yield, which was then compared with the ground-based measurement.

**Property inputs and thermal residue.** Air-atmosphere thermogravimetry (Air-TGA) showed oxidative decomposition near 550–650 °C and a measured terminal residue of 0 wt.% (**Fig. a**). As in the PLAVIS comparison, the measured residual mass fraction was used as a proxy for the ash-mass-fraction descriptor, corresponding to a dimensionless input of 0. The supplied bulk properties and this measured ash proxy were incorporated into the descriptor record used for prediction. The Air-TGA residue was treated as a protocol-dependent proxy rather than as a measurement quantitatively equivalent to RF-plasma ash. The numerical inputs and their sources are provided in Supplementary Note 3.

**Supporting surface characterization.** High-resolution X-ray photoelectron spectroscopy (XPS) revealed a $C–F_x$ component in C 1s and a dominant F 1s peak near 688 eV (**Fig. b**), indicating fluorinated surface chemistry. Following $Ar^+$ etching, the atomic-percent profile became more carbon-rich and showed a marked reduction in the fluorine signal (**Fig. c**). These spectra provided complementary information about the sample's surface chemistry; they were not substituted directly for bulk elemental-composition ratios. Because ion etching can modify polymers, the post-etch composition was not assumed to represent undamaged bulk material.

**Prediction and ground-based comparison.** The equal-weight ensemble, trained only on the 37 MISSE-2 polymers, was applied without retraining. The predicted erosion yield was $1.69 \times 10^{-24}$ $cm^3$ $atom^{-1}$. Ground-based exposure in the RF AO system, with effective fluence determined using a co-exposed Kapton-HN witness, gave a measured erosion yield of $8.82 \times 10^{-25}$ $cm^3$ $atom^{-1}$. The prediction-to-measurement ratio and fold-error were both approximately 1.92, placing the prediction within a factor of two of the measurement (**Fig. e**). No empirical facility multiplier was applied. Although the ensemble prediction was closer to the measurement than the training arithmetic-mean baseline, the training geometric-mean baseline gave a similar fold-error of 1.96.

**Interpretation.** The combination of a reported terminal Air-TGA residue of 0 wt.% and a low measured erosion yield provides a contrast to the residue–resistance trend observed within the PLAVIS grades. This comparison motivates considering material information beyond thermal residue alone when evaluating AO resistance. The XPS observations provide complementary evidence of fluorinated surface chemistry but do not establish fluorination as the cause of the observed resistance. This single-sample comparison illustrates the application of the model to a polymer with undisclosed composition, while the similar geometric-mean baseline error limits claim of ensemble superiority (**Fig. d**).

## Discussion

The present study demonstrates the potential of using readily accessible bulk properties and elemental-composition descriptors for rapid screening of polymer susceptibility to atomic-oxygen erosion. Despite the limited size of the MISSE-2 training set, the model captured meaningful variation in erosion yield and provided predictions of the correct approximate magnitude for independent orbital and ground-based comparisons. However, the external orbital dataset spans a narrow erosion-yield range, and simple constant baselines remained competitive on several metrics. The model should therefore be regarded as a screening framework rather than as a universally accurate predictor of AO durability.

Taken together, the matching predicted and experimental erosion-yield rankings, the grade dependence of the Air-TGA-based ash proxy and density, the Ca- and Si-containing surface species detected on HM and S1, and the SEM-derived morphological trends provide complementary evidence for grade-dependent AO resistance in PLAVIS. Particularly noteworthy is that the Air-TGA residue, which is independent of the experimental erosion-yield calculation, followed the same HM > S1 > ESD > N ordering as AO resistance. These observations are physically consistent with the model assigning relevant predictive information to the ash-related descriptor. Nevertheless, Air-TGA residue is a protocol-dependent proxy rather than a measurement equivalent to RF-plasma ash, and the observed correlations do not establish independent causal effects of individual features, a specific additive-mediated protection mechanism, or general applicability to other polymer families. Density likewise requires cautious interpretation because it enters directly into the calculation of volumetric erosion yield. **Fig. 5a** illustrates the saved model's response to density and ash mass fraction with the remaining descriptors fixed at their training medians. This response slice describes fitted-model sensitivity rather than a causal relationship or a chemically constrained materials-design map. Beyond the PLAVIS comparison, the mystery-sample case in Box 1 provides an additional comparison of the ensemble prediction with the measured erosion yield and training-derived baselines (**Fig. 5b**). The similar performance of the geometric-mean baseline on this specimen limits claims of ensemble superiority.

Several limitations define the current scope of the model, including the small and chemically related training set, selection-conditioned internal validation, and the limited size and range of the external datasets. Future validation should therefore use prospectively fixed descriptor records and predictions, larger and more chemically diverse polymer sets, and simple baselines evaluated under the same protocol (**Fig. 5c**). Matched pre- and post-AO surface characterization and direct chemical analysis of the residual material would further clarify the physical origin of the descriptor–erosion relationships. Such efforts could progressively transform the present screening framework into a more robust and physically interpretable tool for AO-resistant polymer selection.

# Methods

## Datasets and descriptor provenance

The 37-row MISSE-2 training table[4] and nine-row labelled external orbital test set are retained. Training and external values are available in Supplementary Data together with row-level OOF and prediction outputs. PVC is excluded from scoring as a lower-bound observation.[16]

The candidate database contains nine physical-property columns and five elemental-ratio columns. The saved configuration selected eight physical columns plus all five ratios; volume resistivity is not used by the saved model due to the data scaling issue. Inference reads numeric descriptor fields in the saved feature order, not the Formula metadata string.

Note 3. Family-surrogate values are distinguished from measurements. A source tag by itself is not proof of an exact grade-, batch- or condition-matched source.

## Air-atmosphere thermogravimetry

The terminal residual mass fraction was determined by thermogravimetric analysis under an air atmosphere using a Q500 thermogravimetric analyser (TA Instruments) at the Korea Research Institute of Chemical Technology (KRICT). Each sample was heated from room temperature to 900°C at 10°C $min^{-1}$ under an air flow of 100 mL $min^{-1}$.

The residual mass fraction was defined as

$$w_{res} = \frac{m_{res}}{m_0} = \frac{R_{TGA}}{100}$$

where $m_{res}$ is the residual mass at the end of the prescribed TGA procedure, $m_0$ is the initial sample mass, and $R_{TGA}$ is the terminal residue expressed in wt.%. The measured wt.% values were converted to dimensionless mass fractions before being used as model inputs.

**Table 4 | Air-TGA residual masses of the four PLAVIS grades.** Residual mass fraction denotes the terminal Air-TGA residue normalized by the initial sample mass.

| | PLAVIS-N | PLAVIS-ESD | PLAVIS-S1 | PLAVIS-HM |
|---|---|---|---|---|
| Initial mass, $m_0$ (mg) | 31.4220 | 27.2670 | 31.3160 | 36.8070 |
| Residual mass, $m_{res}$ (mg) | 0.06479 | 0.2207 | 0.4213 | 0.7683 |
| Residual mass fraction, $w_{res}$ (%) | 0.2062 | 0.8095 | 1.345 | 2.087 |

Because direct RF-plasma ash measurements were unavailable for the PLAVIS grades, the grade-specific Air-TGA residues were used as proxy inputs for the ash-mass-fraction descriptor in the erosion-yield model (**Table 4**). Oxidative thermogravimetry is commonly used to estimate the non-combustible or inorganic residue remaining after removal of the organic matrix,[19,24] whereas the NASA ash descriptor was obtained by RF air-plasma ashing until no further mass loss was observed.[18] Although the two procedures are not experimentally equivalent, both quantify the residual mass fraction remaining after oxidative removal of the polymeric component.

As a reference for the use of Air-TGA residue as a proxy for the ash-related descriptor, Kapton HN shows a similarly low residue under the two oxidative protocols: 0.3537 wt.% by Air-TGA and 0.441 wt.% in the NASA RF-plasma ashing dataset.[18] Although these values fall within a similar low-residue regime, differences in oxidation protocol and sample provenance preclude treating them as quantitatively equivalent.

## Ground AO exposure and mass-loss calculation

Ground-based AO exposure was performed using an RF oxygen-plasma facility operated with 99.995% $O_2$ at a base pressure below $9 \times 10^{-6}$ Torr, a working pressure of 1 mTorr and an RF power of 60 W. The peak beam energy was 7.875 eV, with a nominal AO flux of $1.42 \times 10^{15}$ atoms $cm^{-2}$ $s^{-1}$. The source-to-sample distance was 10 cm and the exposure duration was 4 h. No additional VUV irradiation or thermal cycling was applied. PLAVIS specimens were 1.1 mm thick with an exposed area of 3 × 3 cm². Before exposure, the specimens were ultrasonically cleaned in deionized water for 90 s and outgassed for more than 48 h at 3 mTorr without bakeout.

A Kapton HN witness with $E_{y,K}$ = 2.81 × $10^{-24}$ $cm^3$ $atom^{-1}$ and $\rho_K$ = 1.42 g $cm^{-3}$ was used to determine the effective AO fluence for each exposure run:

$$F_{\text{eff}} = \frac{\Delta m_K}{\rho_K A_K E_{y,K}}$$

The erosion yield of each sample was then calculated as

$$E_{y,s} = \frac{\Delta m_s}{\rho_s A_s F_{\text{eff}}}$$

The stated nominal flux and duration give 2.0448 × $10^{19}$ atoms $cm^{-2}$. This nominal integral is not a substitute for a measured witness fluence. The recorded sample masses, taken with a balance described as a CAS CAUW-220D with 10 µg readability, are preserved in **Table 5**. Readability is not a measured standard deviation and is not used to manufacture confidence intervals.

**Table 5 | Recorded PLAVIS masses and diagnostic calculation using nominal fluence.** Initial and final masses of PLAVIS specimens before and after AO exposure in a ground AO simulator, corresponding mass losses, and erosion yields calculated using the nominal AO flux.

| | PLAVIS-N | PLAVIS-ESD | PLAVIS-S1 | PLAVIS-HM |
|---|---|---|---|---|
| Initial mass (g) | 1.40457 | 1.33691 | 1.39595 | 1.36168 |
| Final mass (g) | 1.40372 | 1.33628 | 1.39532 | 1.36111 |
| Mass loss (g) | 0.00085 | 0.00063 | 0.00063 | 0.00057 |
| Density (g/$cm^3$) | 1.43 | 1.44 | 1.46 | 1.47 |
| Erosion yield ($cm^3$/atoms·g) | 3.24 × $10^{-24}$ | 2.38 × $10^{-24}$ | 2.35 × $10^{-24}$ | 2.11 × $10^{-24}$ |

## SEM and XPS

SU8600 field-emission scanning electron microscope (Hitachi High-Tech) at a beam voltage of 5.00 kV was used to compare FE-SEM images of pristine PLAVIS-HM with AO-exposed PLAVIS-HM and PLAVIS-N. Samples were coated with osmium before analysis. The supplied SEM panels identify pristine HM and AO-exposed HM/N, with the latter images described as acquired at 30° tilt. The bright-field area plot summarizes top-view images at 30,000×, 50,000×, and 100,000×, and these data were obtained using ImageJ.[25] The image contrast is not a compositional measurement. Original SEM images utilized for image processing are available in Supplementary Figure 1.

XPS analysis was done using a Nexsa G2 surface analysis system (Thermo Fisher Scientific) with a monochromatic Al $K_\alpha$ X-ray source, using an X-ray spot size of 400 μm. We acquired XPS survey spectra for PLAVIS-HM and PLAVIS-S1. The XPS panels show survey spectra for HM and S1. For the mystery sample (MS), both survey and high-resolution C 1s, F 1s, O 1s and N 1s spectra were acquired. High-resolution spectra were analysed using Avantage software (Thermo Fisher Scientific). Depth-dependent elemental composition was additionally examined by two consecutive 120-s $Ar^+$ etching cycles.

FE-SEM imaging and XPS data acquisition were performed at KAIST Analysis Center for Research Advancement (KARA), whereas subsequent image analysis and spectral fitting were carried out by the authors.

## Archived model and validation protocol

The saved model metadata specify scikit-learn[26]1.5.1 and XGBoost[15] 2.1.4. Predictions were reproduced in the ISMRC_3 environment with these versions. Every base estimator is a *TransformedTargetRegressor* with $\log_{10}(y)$ target and inverse $10^y$ (epsilon zero). Median imputation and standardization precede each estimator. The ratio block is retained, and the top eight physical features were selected by absolute correlation with the original-scale target on the complete training table before the outer validation splits. This selection step was not nested in LOOCV.

RidgeCV selects alpha from its stored grid using five-fold internal validation; its enclosing preprocessing is fitted on the outer training fold, not separately inside RidgeCV's own internal splits. GPR uses ConstantKernel × RBF + WhiteKernel with target normalization. The two XGBoost learners use hyperparameter searches within the archived outer folds. Their GridSearchCV scoring is RMSE in the original target units, even though the estimators fit a log target. The saved final random-forest-style model has 240 trees, maximum depth 10 and subsample 0.7; the sequential gradient booster has 400 trees, depth 1 and learning rate 0.05. Full saved parameters and the source code remain the authoritative model specification.

Four blending schemes were compared through additional cross-validation over the full training table and the equal scheme was selected using a one-standard-error rule. Because feature and scheme decisions were not independently repeated inside every outer LOOCV split, this procedure does not justify the former claim that all model-development choices were fully nested. The saved model is preserved; no revised model is substituted for its reported external predictions.

Feature importance is taken from the saved Ridge coefficients and the supplied ensemble permutation results. The code performs 20 permutations per feature on the fitted training data. This is descriptive training-set importance, not a held-out estimate of a feature's incremental value.

## Retrospective baseline and sensitivity analyses

Arithmetic and geometric means are computed from all 37 orbital training labels for external comparisons; their internal LOO versions exclude the test row. Saved RidgeCV, GPR and both tree predictions are compared with the same ensemble on identical sample/measurement pairs. New Ridge analyses use six fixed feature groups: all 14 curated inputs, composition only, physical properties only, all except ash, density plus ash, and ash only. Unlike the saved top-k procedure, these groups are not chosen by correlations with the target. Alpha is selected from $10^{-4}$ through $10^{4}$ in nine logarithmically spaced steps, using five-fold shuffled inner validation on log-target mean-squared error. Imputation and scaling are inside that search pipeline. Outer validation is leave-one-out with inner seed 42 + outer-fold index; final training-only refits use seed 42. All group results are reported and no external score is used to choose a winning feature group for the primary model.

## Evaluation metrics

Fold-error is max(predicted/measured, measured/predicted), which is always at least one. Pred/Meas can be below one and must not be called a fold-error in that case. $R^2$(log) and $R^2$(linear) are computed separately; RMSE and MAE are reported in $cm^3$ $atom^{-1}$, with log-RMSE also preserved in the machine-readable output. Metrics for the two provisional ground measurement records are computed separately for PLAVIS alone and for all five films. Archived OOF and external-orbital CSVs are limited to their stored precision; these are not claimed to contain the original unrounded instrument measurements. Reproduced experimental predictions are retained at full floating-point output precision.

Paired bootstrap intervals use 20,000 resamples with replacement and seed 42. They are conditional on the supplied predictions and labels, do not retrain the model and do not account for measurement uncertainty. No calibrated specimen-level prediction intervals or experimentally verified classification threshold is claimed.

# Code availability

The full training and evaluation pipeline (CLI entry points, source modules `data.py`, `models.py`, `train_eval.py`, `viz.py`, and the inference script `predict.py`) is released under an open-source license at https://github.com/YoungwooChoiKAIST/AO_prediction. The command *python scripts/train_and_evaluate.py --dataset training.csv --testset "validation(post-MISSE-2).csv" --no-merge_testset* reproduces the training-set cross-validation and the post-MISSE-2 validation comparison, and *python scripts/predict.py --input "validation(experimental).csv"* reproduces the experimental-set predictions.

# Data availability

The training table (training.csv), validation descriptor tables, per-row out-of-fold predictions, external-validation predictions and evaluation metrics are to be provided as Supplementary Data.

Polymer-level training labels were taken from the NASA MISSE-2 PEACE reports,[4,8] with post-MISSE-2 labels from subsequent campaigns. Bulk properties were collated from MatWeb and PolyInfo. Measured erosion yields for the four PLAVIS grades are reported in Table 3.

## Supporting Information

Supplementary Information is available for this paper, comprising Supplementary Notes 1 and 2, Supplementary Table 1, Supplementary Figures 1-3, and Supplementary References 1–28. Supplementary Data 1 (harmonized training and validation property tables, per-row out-of-fold predictions, and external-validation predictions referenced in Data availability) accompanies this paper.

## Acknowledgements

This work was supported by the National Research Foundation of Korea (NRF) through grants funded by the Korean government (MSIT) (Grant Nos. RS-2023-00247245 and RS-2025-02213804). The latter grant supported the In-Space Servicing and Manufacturing Research Center in 2026.

## Author contributions

Y.C. implemented and trained the machine-learning models, generated predictions and evaluated model performance. D.K. conducted atomic oxygen erosion experiments using a space-environment simulator and performed materials characterization, including scanning electron microscopy, thermogravimetric analysis and X-ray photoelectron spectroscopy. S.H. and H.J.K. jointly supervised the study. S.H. provided guidance on machine-learning methodology and model evaluation. H.J.K. conceptualized the study, designed the experimental framework and provided expertise in space materials and guidance on the interpretation of atomic oxygen erosion mechanisms. Y.C. and D.K. contributed equally to writing the manuscript.

# Figures

**Figure 1 | AO exposure, training data and saved modelling workflow.**

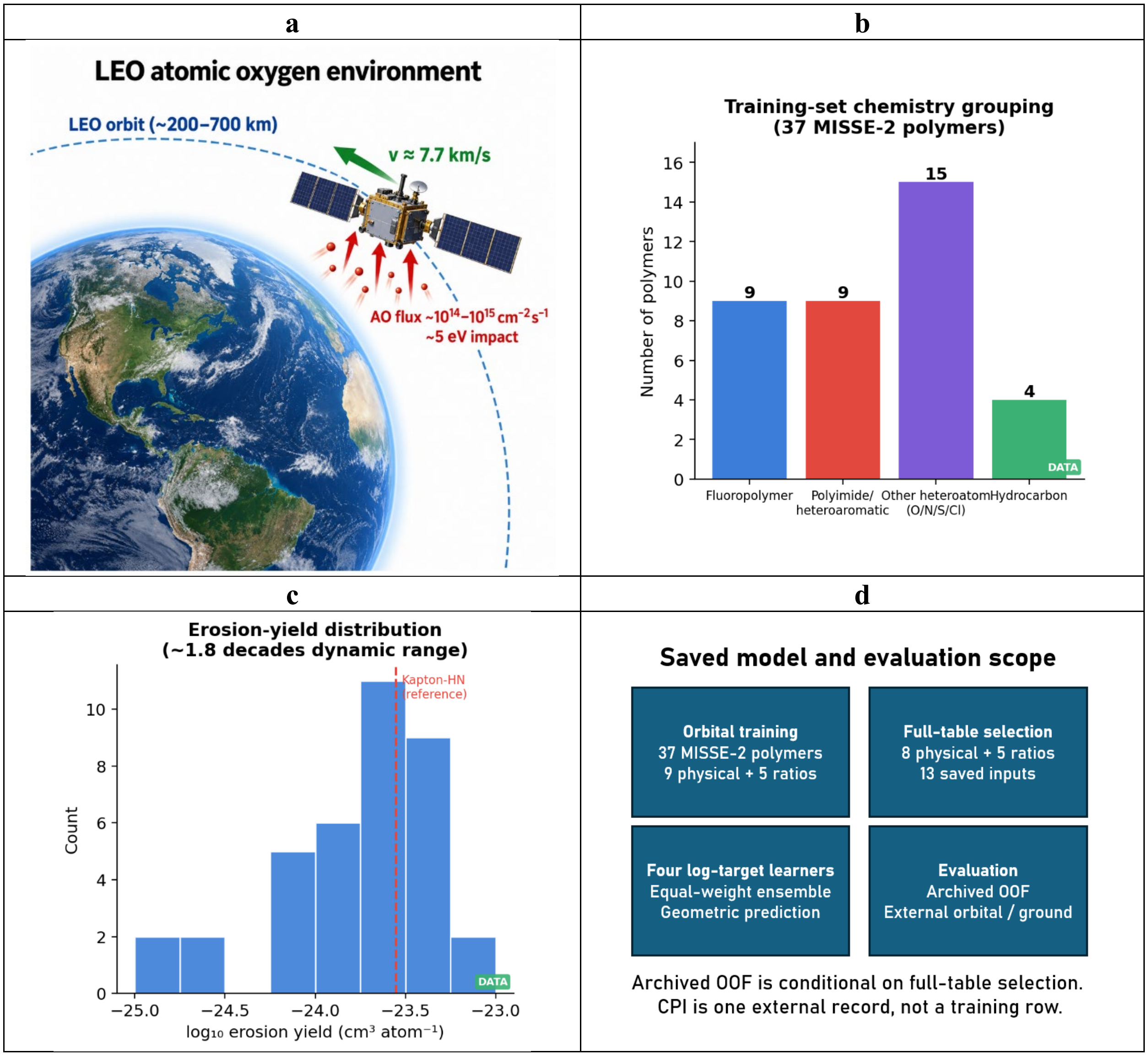

a, Original AO exposure schematic. b,c, Original training-family counts and target distribution. d, Workflow showing eight selected bulk inputs plus five elemental ratios. The top-k feature choice precedes the archived outer folds, so internal performance is conditional on that selection.

**Figure 2 | Descriptor associations in the saved model.**

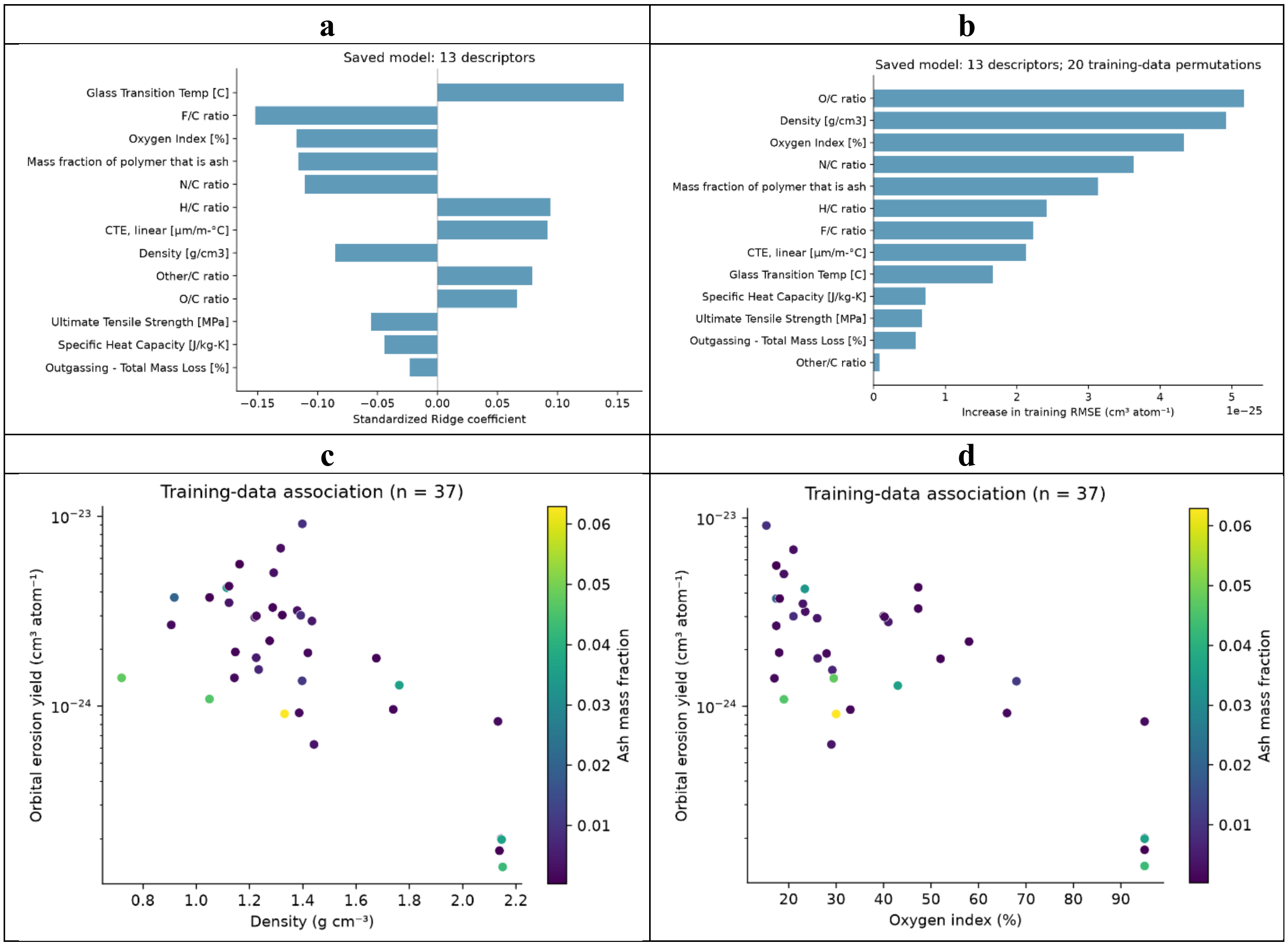

a, Actual standardized Ridge coefficients for 13 inputs; Tg has the largest absolute coefficient. b, Training-data permutation importance, 20 repeats, measuring change in original-unit RMSE. Neither analysis establishes causality or external feature benefit. c,d, Training erosion yield versus density and oxygen index, respectively, with ash as the colour scale. No monotonic trend is imposed.

**Figure 3 | Consistently recalculated performance.**

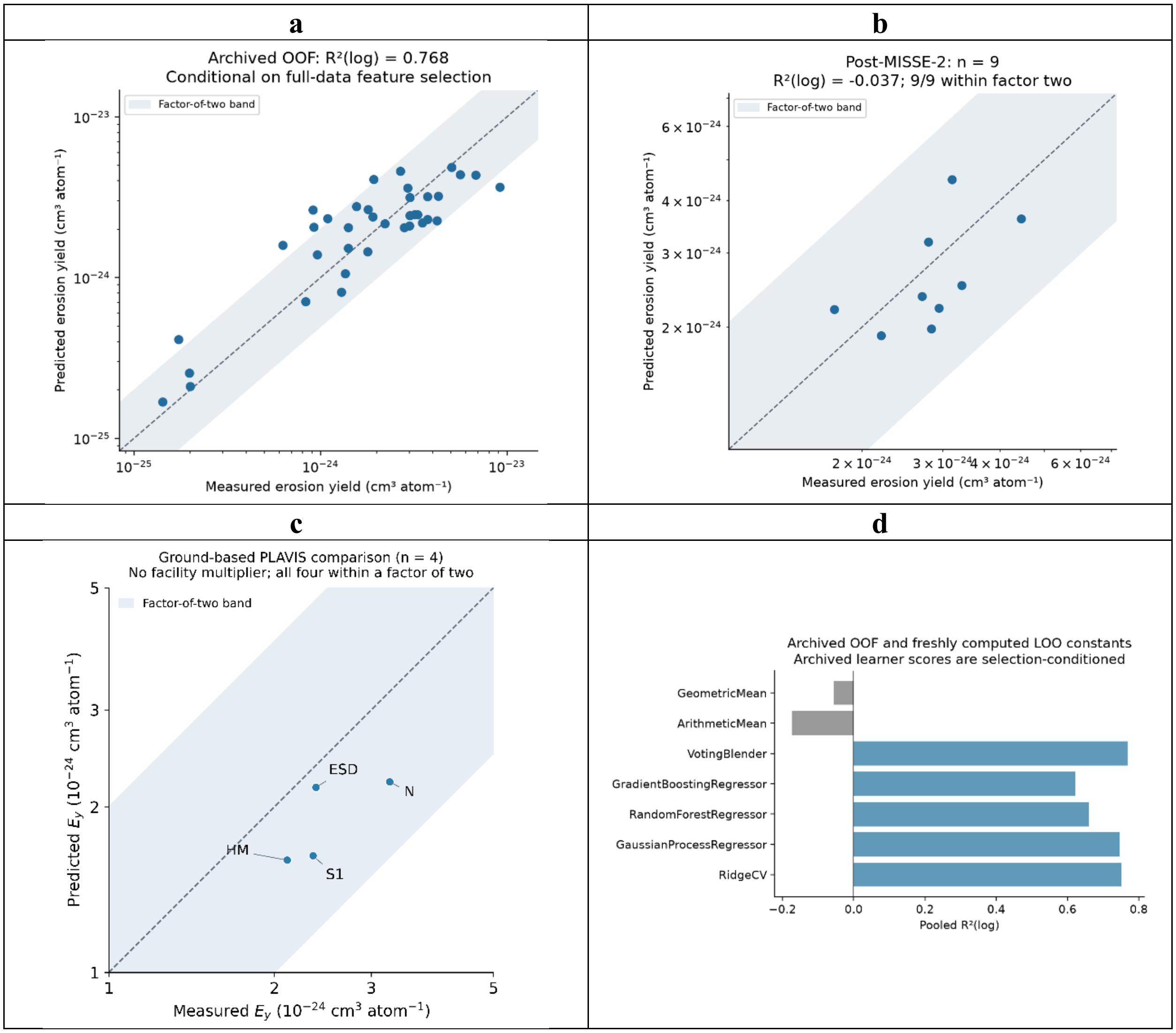

a, Archived ensemble OOF, $R^2$(log) = 0.768, conditional on selection using the full training table. b, Nine post-MISSE-2 pairs, $R^2$(log) = −0.037; PVC excluded. c, Predicted versus measured erosion yields for the four PLAVIS grades (N, ESD, S1 and HM), using the values displayed in Table 3. The dashed line denotes equality and the shaded region indicates predictions within a factor of two of measurement, not a confidence interval. All four predictions fall within this band; no facility multiplier is applied.

**Figure 4 | PLAVIS thermal residue and surface characterization.**

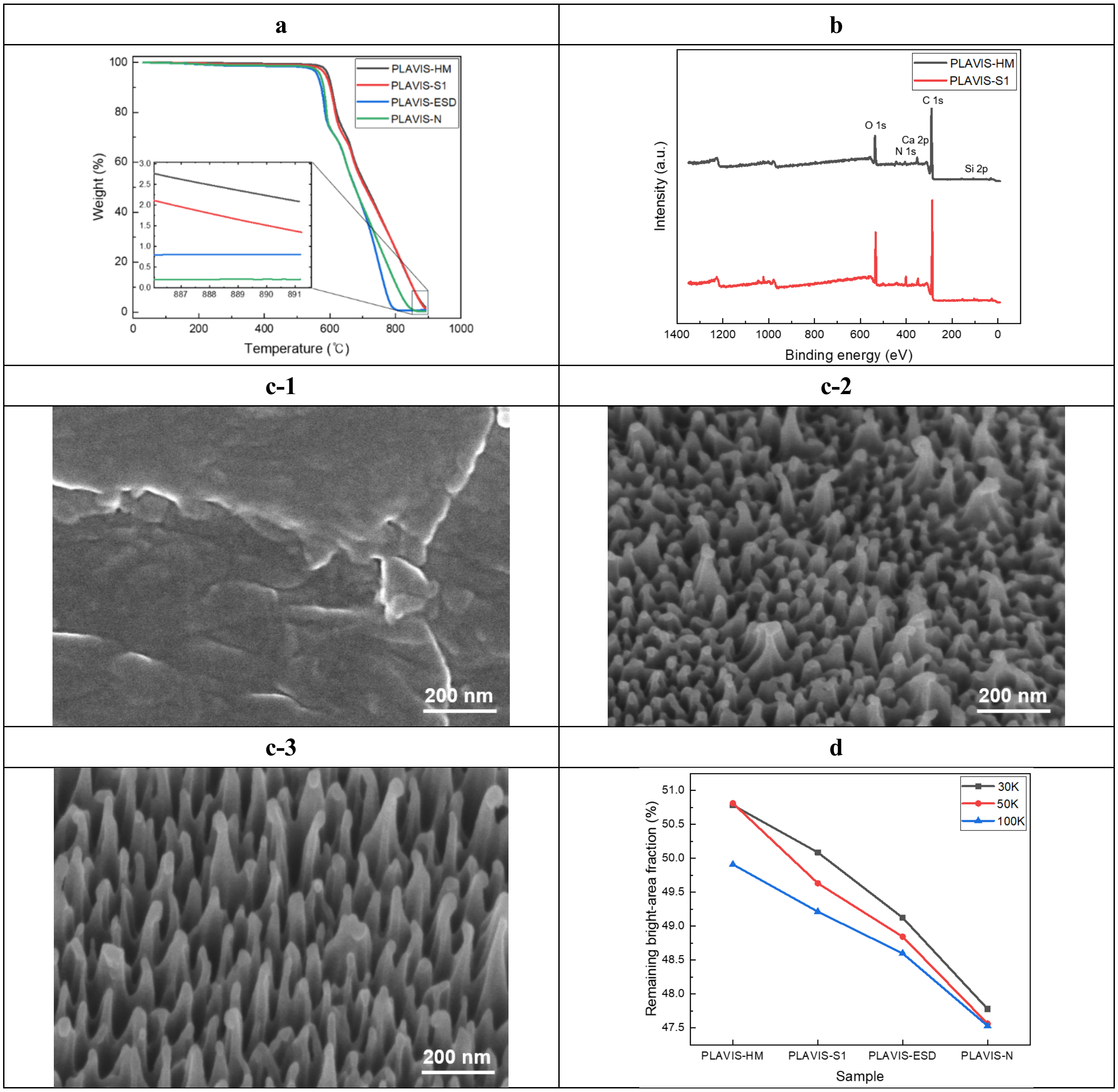

a, Air-TGA with terminal inset; endpoints need not be equilibrium ash. b, XPS surveys with assigned Ca/Si features; chemical phases and AO enrichment are unproven. c-1, Pristine HM; c-2, exposed HM; c-3, exposed N; exposed panels are reported at 30° tilt, scale bars 200 nm. d, Bright-area fraction (%) from the reported top-view images. The original panel's word "remaining" denotes thresholded bright image area, not retained material mass. Magnifications are not independent specimen replicates. Comparison with erosion ordering is conditional on reconciliation of the ground labels.

**Box 1 Figure | Thermal residue and surface chemistry of the held-out mystery sample.**

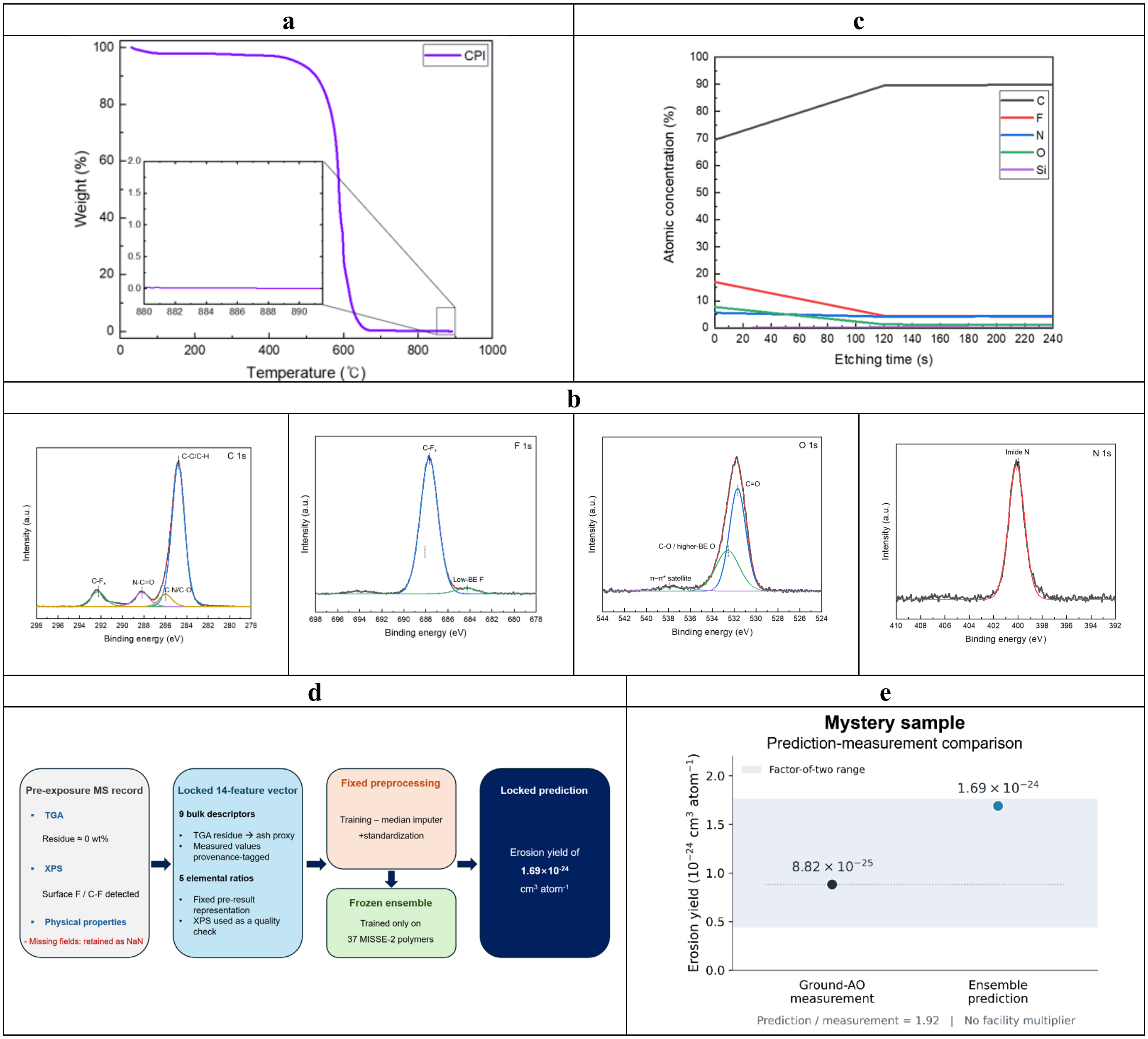

**a,** Thermogravimetric analysis (TGA) of the mystery sample (MS) performed in an air atmosphere. **b,** High-resolution XPS spectra of MS surface with peak deconvolution: (b-1) C 1s, (b-2) F 1s, (b-3) O 1s, and (b-4) N 1s spectra. **c,** Atomic concentrations of the constituent elements in MS as a function of $Ar^+$ monatomic ion etching time. XPS elemental compositions were measured before etching and after two consecutive 120 s etching cycles. **d,** Experimental evidence is mapped to the frozen 14-feature vector. Missing fields remain NaN and undergo the preprocessing fixed during MISSE-2 model development. **e,** Comparison of the reported ground-based AO erosion yield of MS ($8.82 \times 10^{-25}$ $cm^3$ $atom^{-1}$) with the saved ensemble prediction ($1.69 \times 10^{-24}$ $cm^3$ $atom^{-1}$). The prediction-to-measurement ratio and fold-error are both approximately 1.92. Shading denotes a factor-of-two range around the measurement, not a confidence interval. No empirical facility multiplier is applied.

**Figure 5 | Model response and transparent screening assessment.**

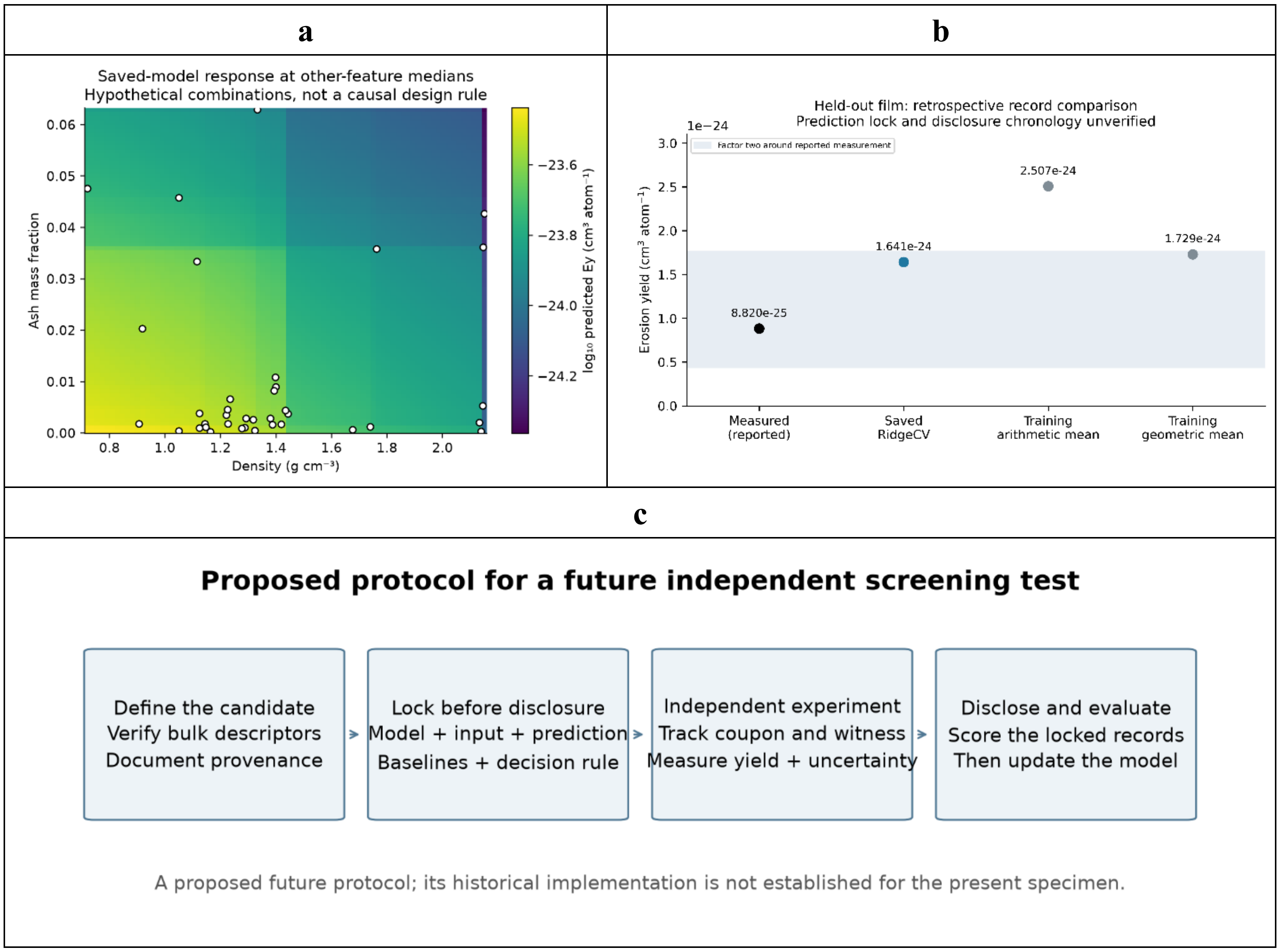

a, A density–ash response slice from the saved 13-input model with all other features fixed at training medians. The overlaid points mark observed training density/ash pairs. The slice is a hypothetical response of the model, not a measured or chemically constrained design map. b, Reported erosion yield of the mystery sample compared with the saved RidgeCV prediction and arithmetic- and geometric-mean baselines calculated from the MISSE-2 training labels. Shading indicates a factor-of-two range around the reported measurement, not a confidence interval. Prediction-lock and disclosure chronology remain unverified; no facility multiplier is applied. c, A proposed future workflow: fix records and predictions, independently disclose measurements, evaluate, and only then update the model.